\documentclass[useAMS,usenatbib]{mn2e}  

\usepackage{amssymb,amsmath}
\usepackage{amsfonts}
\usepackage{graphicx}
\usepackage{tabularx}
\usepackage{color}

\numberwithin{equation}{section}

\def\d{{\rm d}}
\def\e{{\rm e}}
\newcommand{\ee}{\end{equation}}
\newcommand{\be}{\begin{equation}}
\newcommand{\p}{\partial}

\newcommand{\case}{\textstyle\frac}

\begin{document}

\title[Effect of ILR on Spiral Density Waves]{Effect of Inner Lindblad Resonance on Spiral Density Waves Propagation in Disc Galaxies:  Reflection over Absorption}

\author[Polyachenko \& Shukhman]
       {E.~V.~Polyachenko,$^1$\thanks{E-mail: epolyach@inasan.ru}
        I. G. Shukhman,$^2$\thanks{E-mail: shukhman@iszf.irk.ru}\\
       $^1$Institute of Astronomy, Russian Academy of Sciences, 48 Pyatnitskya St., Moscow 119017, Russia\\
       $^2$Institute of Solar-Terrestrial Physics, Russian Academy of Sciences, Siberian Branch, P.O. Box 291, Irkutsk 664033, Russia}



\maketitle

\begin{abstract}

Interaction of spiral density waves with stars in the vicinity of the inner Lindblad resonance in galactic discs is investigated using the linear perturbation theory and the leading orders in the epicyclic and WKB approximations. In analogy with shear flows in hydrodynamics, we conjecture that a weak nonlinearity in a narrow resonance region modifies the standard (Landau\,-\,Lin) bypass rule of the singularity to the integration in the principal value sense.
This indeed leads to the reflection of the spiral wave instead of absorption, but the detailed picture looks awkward: the intervals of the wave weakening alternate with the intervals of the wave growth, so that the net absorption is absent. Incidentally, we rectify the result concerning leading spiral waves obtained earlier for the standard bypass rule.
\end{abstract}


\section{Introduction}

A core of any theory of spiral structures is a mechanism for spiral formation and support \citep{BT2008, B14, DB14}. One of them suggests amplification of the density waves through over\,-\,reflection, when the amplitude of the reflected wave is higher than of the incident wave. Over\,-\,reflection occurs at the corotation resonance (CR) because it separates the inner disc with the negative wave angular momentum density from the outer disc where this density is positive \citep{Lau76}. The amplifier can give rise to a global unstable pattern, when the outer wave departs freely taking out positive angular momentum, and the inner wave is looped continuously reflecting from the centre or a barrier \citep{Mark77, Toomre81}.

\begin{figure*}
 \begin{center}
 \includegraphics[width=170mm]{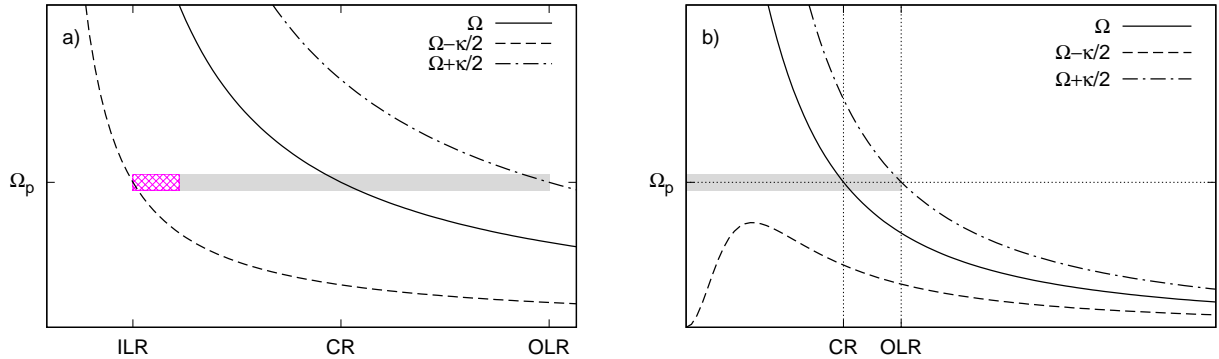}
 \caption{\protect\footnotesize -- The spiral pattern speed $\Omega_{\rm p}$, frequency curves $\Omega(R)$ and $\Omega \pm \varkappa(R)/2$ and principal resonances for cuspy (a) and cored (b) galactic models. The designations are given in Sect. 2. The principal ranges are shown by gray thick lines; the magenta pattern on a) shows the resonance region of interest. }
 \label{Fig.0}
 \end{center}
\end{figure*}

Lindblad resonances are known as perfect absorbers of density waves, which propagate in a so-called principal range \citep{LS66}, i.e., between the inner Lindblad resonance (ILR) and the outer Lindblad resonance (OLR), see Fig.\,\ref{Fig.0}a. In the outer disc, OLR helps to absorb the positive angular momentum density wave with no reflection. Near ILR, the trailing spiral waves have the group velocity directed inwards and the leading waves have the group velocity directed outwards, and each wave decays in the direction of its propagation (Mark 1974). This evidently breaks the feedback loop, and is believed to be a reason for hampering the pattern growth. A way to avoid ILR is to decrease a central mass concentration and increase a thin disc mass: in this case the spiral pattern speed will probably exceed a maximum value at which ILR is possible (Fig.\,\ref{Fig.0}b). Another way is to hide ILR inside a region opaque to spiral waves, which would reflect them back to CR \citep{T77, LauBertin78}. If nevertheless the feedback loop in the centre is interrupted, only short-lived (transient) spiral waves are possible \citep{Toomre81}.

Fluid discs are often used to describe wave dynamics in (stellar) galactic discs \citep{LinLau75, LauBertin78, LinLau79, Pannatoni1983}. Similarly to the mentioned above, CR separates positive and negative wave angular momentum density and thus works as an amplifier. But interaction of waves with Lindblad resonances is totally different. Without selfgravity, the fluid wave can go freely through the resonances \citep{LFK94}, and accounting for selfgravity will hardly change things. Some authors believe that there is some dissipation and attenuation of the transmitted wave, but this belief mainly goes back to unpublished works by Lau in mid 1970-th, and no rigorous results available now \citep[][G. Bertin, 2018, private communication]{B14}. Thus serious efforts are needed to mimic the behaviour of stellar disc using its fluid counterpart.

Turning back to stellar discs, many authors emphasise that if ILR is saturated, it can be changed from the perfect absorber to a reflector \citep{Lin70, Contopoulos70a, Contopoulos70b, LauLinMark76, Toomre81}. This question, however, requires a departure from the purely linear formulation of the problem. The main goal of the paper is to explore this opportunity. We still consider a linear problem in general, but we assume that the nonlinearity affects a narrow region near the resonance and changes a so-called bypass rule for traversing the pole. This bypass rule lies in the foundation of Mark's derivation (1971, 1974) of the perfect absorption of density waves on Lindblad resonances.

Using the analogy with the problems of wave behaviour in fluid shear flows, we {\it conjecture} that the nonlinearity modifies the standard Lin\,-\,Landau bypass rule, and dictates calculation of the integral in the principal value sense. We show that in this case, similar to the analogous problem on Rossby waves in fluids, perfect reflection instead of perfect absorption of waves occurs.

The structure of the paper is as follows. In the next section we give an extended review of bypass rules applicable in collisionless stellar dynamics (or collisionless plasma) and in hydrodynamics. Section 3 contains the derivation of the dispersion relation with the standard bypass rule and discusses a difference between our and Mark's results. Chapter 4 contains the modification of the bypass rule, which presumably occurs when the spiral wave or bar is already formed. A discussion of the obtained results in compare with the well-known work by \cite{LBK72} is given in Chapter 5. The concluding section summarises our results and relates them with current problems of bar formation. Appendix A contains the derivation of the resonance contribution to Lin\,-\,Shu reduction factor using action\,-\,angle technique. Appendix B gives the asymptotic form of dispersion relations for short-wave spirals for two possible cases of bypass rule (it is instructive because it allows to avoid mistakes when deriving the asymptotic form of dispersion relations for very short waves starting from the dispersion relation of general form containing the modified Bessel function).

\section{Bypass rules in theories of stellar and fluid media}

Bright spiral arms are composed of newly formed stars and thus trace star formation regions rich of gas. The gas dynamics is responsive to underlying mass dominating stellar population usually considered as a fixed axially symmetric disc with perturbations in the form of spiral density waves. The propagation of density waves is described by the dispersion relation
\begin{equation}
	\frac{k_T}{k}\,s_k ={\cal F}^{(\rm LS)}(\nu,x) \ ,
	\label{eq:ep1_lsk}
\end{equation}
where $k = k(r)$ is the radial wavenumber of the perturbed gravitational potential and surface density:
\begin{equation}
\left[\begin{array}{l}
\!\!\!\delta\Phi(r,\varphi,t)\\
\!\!\!\delta\sigma(r,\varphi,t)
\end{array}\!\!\!\right]
=
\left[\begin{array}
{c}
\!\!\!V(r)\\
\!\!\!S(r)
\end{array}\!\!\!
\right]\,
\exp\Bigl\{i\, \Bigl[\int\limits^r k(r')\,\d r' + m\varphi - \omega t\Bigr]\Bigr\},
\label{eq:WKB}
\end{equation}
$k_T(r) \equiv \varkappa^2(r)/(2\pi G\, \sigma_0(r))$, $s_k \equiv \mathrm{sgn}\,[{\rm Re} (k)]$,  $\nu\equiv (\omega- m\Omega)/\varkappa$ is the dimensionless frequency, $\omega$ is the frequency of disturbance (supposed to be real since stationary problem is considered), $\Omega(r)$ is the angular velocity, $\varkappa\,(r)$ is the epicyclic frequency, $m=2$ is the (azimuthal) number of spiral arms, $x \equiv k^2\,c^2/\varkappa^2$, $\sigma_0(r)$ is the equilibrium surface density of the stellar disc, and $c\,(r)$  is the radial velocity dispersion. Function ${\cal F}^{(\rm LS)}$ is a so-called Lin\,-\,Shu reduction factor, which is \footnote{Originally, the Lin\,-\,Shu reduction factor is our $(1-\nu^2) {\cal F}^{(\rm LS)}$.}:
\begin{equation}
  {\cal F}^{(\rm LS)}(\nu,x)= \frac{e^{-x}}{x}
\sum\limits_{-\infty}^{\infty}\frac{n\,I_n(x)}{n-\nu}
  \label{eq:ep1_rfls}
\end{equation}
for the Schwarzschild distribution function (DF),
\begin{equation}
F(E,R)=\frac{\gamma\,\sigma_0(R)}{2\pi c^2(R)}\,\exp \left[
-\frac{E-E_{\rm c}(R)}{c^2(R)}\right]\, .
 \label{eq:DF}
\end{equation}
Here $I_n(z)$ are the modified Bessel functions of the first kind, $R$ denotes the radius of the guiding centre of the epicycle, which is connected to the angular momentum of the star
$L=\Omega(R)\,R^2$; $E=\frac{1}{2}\,(v_r^2+v_{\varphi}^2)+\Phi_0(r)$ is the star's energy; $E_{\rm c}(R) \equiv \frac{1}{2}\,\Omega^2(R)\,R^2+\Phi_0(R)$ is the energy of circular motion of the guiding centre, so that difference $E-E_{\rm c}(R)$ is the energy of epicyclic motion, $E-E_{\rm c}(R)=\frac{1}{2}\,(v_r^2+\gamma^2{\tilde v}_{\varphi}^2)$; $\gamma=2\Omega(R)/\varkappa(R)$, ${\tilde v}_{\varphi}=v_{\varphi}-\Omega(r)\,r$. Note also that $\gamma\,{\tilde v}_{\varphi} \equiv \varkappa\,(R-r)$ \citep[see, e.g.,][]{BT2008}.

It is worth recalling the conditions used to obtain eq.~(\ref{eq:ep1_lsk}): (i) $R$ and $r$ are far from resonances, (ii) the leading orders of WKB and epicyclic approximations are taken into account only, and (iii) two small parameters of the problem, $1/(kr)$ and $\epsilon \equiv a/r$, where $a\equiv c/\varkappa$ is the epicycle size, are of the same order: $(kr)^{-1}\sim\epsilon$. The latter means that the amplitudes of the potential $V$ and of the surface density $S$, defined by the expression (\ref{eq:WKB}), are assumed to be virtually unchanged within the wavelength and the size of the epicycle.

The dispersion relation (\ref{eq:ep1_lsk}) gives the dependence of the real wavenumber $k(r)$ on radius $r$ for various values ​​of the spiral pattern speed $\Omega_{\rm p} \equiv \omega/m $. Positive and negative values ​​of $k$ correspond to the trailing and leading spirals, respectively. In agreement with the ``antispiral'' theorem \citep{LBO67}, relation (\ref{eq:ep1_lsk}) does not distinguish the type of the spiral, since for a given $\Omega_{\rm p}$ both solutions, $k=\pm |k(r,\Omega_{\rm p})|$, exist.

We go on to consider the generalisation of the dispersion relation (\ref{eq:ep1_lsk}) valid near ILR. The wavenumber $k$ now allowed to be complex, but the imaginary part $k_i$ is assumed to be much smaller than the real part $k_r$, and vanishing outside the resonance region.\footnote{Amplitudes $V(r)$ and $S(r)$ vary slowly on $r$-coordinate due to the medium inhomogeneity. In principle, this variation can also be described by introduction of the imaginary part of the wavenumber, $k_i=\d\ln|V|/\d r$. However, this ``purely kinematic'' component should be distinguished from one due to dissipative processes. Besides, it has the next order of magnitude in $\epsilon$, and thus can be neglected in the leading order.}

For the first time, the problem of the wave interaction with ILR under the above mentioned conditions was considered by \cite{Mark71}. The next paper \citep{Mark74} carries out calculations involving the next order in $\epsilon $ \citep[similar to][]{Shu70}, but here we restrict ourselves to the leading order.

The resonance contribution ${\cal F}^{\rm RES}$ to the reduction factor, ${\cal F}={\cal F}^{\rm NR} + {\cal F}^{\rm RES}$, corresponding in the sum (\ref{eq:ep1_rfls}) to the term containing denominator $\nu(R)+1$ (that is, $n=-1$), can be obtained in the form:
\begin{multline}
{\cal F}^{\rm RES}(y,u_L)= -\frac{i\,\ell}{r\,\epsilon\,x}\\
\times\int\limits_{-\infty}^0\,I_1(x-\mu\,y)\,\,e^{-(x-\mu\,y)}
\,e^{-\frac{1}{2}\,\mu^2-i\,u_L\,\mu}\,\d\mu.
\label{eq:RF_Mark_i}
\end{multline}
Here the variable $u_L \equiv (r-R_L)/(r\epsilon)$ characterises distance to the resonance measured in units of the epicyclic radius $a$,  $\ell\approx [(\d\nu/\d R)_{R=R_L}]^{-1}$ and $y=k c/\varkappa$, $x=y^2$. In Appendix A, we give a new derivation of eq. (\ref{eq:RF_Mark_i}) using the action\,-\,angle variables, rather than Mark's, who used the standard Lin\,-\,Shu technique \citep[see also Appendix K of][]{BT2008}.
\smallskip

Derivation of (\ref{eq:RF_Mark_i}) essentially uses a so-called bypass rule following from the causality principle. It appears in the density response calculation $\delta\sigma$ as one needs to take the integral over velocities from the perturbed distribution function $\delta f$ having  a pole. Indeed, in new variables
$$
  \xi \equiv \frac{v_r}{R\,\varkappa(R)}, \ \ \eta \equiv 1-\frac{r}{R},
$$
the equilibrium DF (\ref{eq:DF})
\begin{equation}
  F=P(R)\,\exp\Bigl[-\frac{v_r^2+\varkappa^2(R)\,(r-R)^2}{2\,c^2(R)}\Bigr]
\end{equation}
can be written as
\begin{equation}
  F=P(R)\,\exp\Bigl[-\frac{\xi^2+\eta^2}{2\,(c/\varkappa\,R)^2}\Bigr],
\end{equation}
and the resonant denominator $[\nu(R)+1]^{-1}$ in expression (\ref{eq:delta_f_res}) for $\delta\!f^{\rm RES}$ near $r=R_L$ becomes
\begin{equation}
	\frac{1}{\nu(R)+1}\approx \frac{\ell}{r}\,\frac{1}{\eta-\eta_L},
	\label{eq:ep1_trick}
\end{equation}
where
$\eta_L=1-r/R_L$, i.e. $\delta\!f^{\rm RES}$ contains a pole at $\eta=\eta_L$.

To overcome the ambiguity of the integral, the causality principle leading to a bypass rule of the pole is used. In the linear theory, this principle suggests no disturbance in the distant past. Assuming the perturbations $\sim e^{-i\omega t}$, this formally corresponds to addition of a small {\it positive} imaginary part to real $\omega$. Then the term $1/[\Omega (z)-\omega]$ near the resonance point $z=z_c$ obeying
$\Omega(z_c)=\omega$ can be transformed to:
\begin{equation}
	\frac{1}{\Omega(z_c)+ \Omega'_c\,(z-z_c)-(\omega+i\delta)}=
	\frac{1}{\Omega'_c}\,\frac{1}{z-(z_c+i\delta/\Omega'_c)},
\end{equation}
where $\delta\to +0$. Therefore, the bypass of the pole in complex plane $z$ must be done from below for $\Omega'_c>0$, and from above for $\Omega'_c <0$. This rule is called the Landau bypass rule, and formally follows from the solution of the initial problem by the Laplace transform method \citep{Landau46}. Applied to real galaxies with $\nu'\equiv [\d\nu/\d R]_{R=R_L}>0$, the rule implies the bypass of the pole $\eta=\eta_c$ {\it from above}.

A similar problem with singularity arises in hydrodynamics when addressing to inviscid version of the Navier-Stokes equation, that is, the Rayleigh equation. In a plane-parallel shear flow with the velocity $U_x=u(y)$, the singular point for a stationary wave of the form $\sim\exp[i\,(kx-\omega\, t)] = \exp[i\,k(x-c\,t)]$ appears at $y=y_c$, $u(y_c)=c$, and is called {\it critical level}. The problem was solved by Lin \citep[see, e.g., ][]{Lin55}, but this time by assuming negligible viscosity rather than considering the initial value problem. It turned out that the bypass rule in integrals of the form $\int \d y\,(...)/[u(y)-c]$ coincides with the Landau bypass rule, and thus it is now known as the Landau\,-\,Lin bypass rule.

In the theory of shear flows, the bypass problem is formulated in terms of a so-called {\it phase change} of the logarithm in the solution $\psi(y)$ of the Rayleigh equation, as one crosses the critical level along $y$-axis. The general solution of the Rayleigh equation near the singular point is
$$
\psi(y)=A\,f_A(\zeta)+B\,f_B(\zeta), \ \ \ \zeta\equiv y-y_c,
$$
where
$$
	f_A(\zeta)= 1+q_c\,\zeta\,\ln\zeta,\ \  f_B=\zeta+...,
$$
and the problem is how to understand the logarithm for $\zeta<0$, as one passes across the critical level. If we write
$$
\ln \zeta = \ln| \zeta |+i\,\Phi
$$
and assume that for $\zeta>0$ the phase of the logarithm $\Phi$ is zero, then for $\zeta<0 $ the phase $\Phi$ turns out to be $-\pi$, because the cut in the complex plane $y$ must be drawn upward from the branch point of the logarithm, provided $u'_c>0$. Thus, the phase change of the logarithm  $\Phi$ in the transition across the critical level from $\zeta>0$ to $\zeta<0$ is equal to $-\pi$, which means the bypass rule from below.

The Landau\,-\,Lin rule, however, is not the only bypass rule, and there can be more subtle situations. For the first time it was shown by \cite{BB69} and independently by \cite{Davis69}, that weak nonlinearity leads to the formation of a narrow layer in the neighbourhood of a critical level of width $l_N\sim\varepsilon^{1/2}$, where $\varepsilon$ is the measure of nonlinearity (wave amplitude). This is a so-called {\it nonlinear} critical layer (CL). This nonlinear CL should be distinguished from a linear viscous CL, width of which is determined by the viscosity ${\bar \nu}$ and is equal to $l_{\nu}\sim {\bar \nu}^{1/3}$. The viscous CL exists even in linear problems, where the amplitude $\varepsilon$ is the smallest parameter, i.e. even smaller than the viscosity (in the corresponding dimensionless units), which is also very small. For $\varepsilon>{\bar\nu}^{2/3}$, nonlinearity is more important than viscosity.\footnote{This is true provided the wave is stationary. In the case of instability with a finite growth rate $\gamma_L$, the resonance can be neglected. However if growth rate is very small, i.e. near the instability threshold, the growth rate (nonstationarity) determines its own critical layer with width $l_t\sim\gamma_L \ll 1 $, and then three factors compete, \citep[see][for more details]{CS96a,CS96b}.}

Although outside the CLs viscosity and nonlinearity do not affect the flow, in case of dominating nonlinearity the phase change of the logarithm instead of $\Phi=-\pi$ is equal to zero. The latter means that an integral with a pole type singularity must be calculated as an integral in the principal value sense. In general, the phase change $\Phi$ can take any value in the interval between $-\pi$ and zero, depending on the value of the parameter $\lambda\equiv{\bar\nu}/\varepsilon^{3/2}$ characterising a ratio between the weak viscosity ${\bar\nu}$ and the weak nonlinearity $\varepsilon$ \citep{Haberman72}.

\section{Dispersion relation for the Landau\,-\,Lin bypass rule}

This section considers the standard (Landau\,-\,Lin) bypass rule used in \citet{Mark71, Mark74}. The resonance contribution to the right-hand side (r.h.s.) ${\cal F}^{\rm RES}_{\frown}$ of the dispersion relation
  \be
\frac{k_T}{k}\,s_k={\cal F}^{\rm RES}_{\frown}+{\cal F}^{\rm NR}
\label{eq:stand1}
\ee
is given by the expression:
\begin{multline}
{\cal F}^{\rm RES}_{\frown}(y,u_L)= -\frac{i\,\ell}{r\,\epsilon\,x}\\
\times\int\limits_{-\infty}^0\,I_1(x-\mu\,y)\,\,e^{-(x-\mu\,y)}
\,e^{-\frac{1}{2}\,\mu^2-i\,u_L\,\mu}\,\d\mu
\label{eq: RF_Mark_a}
\end{multline}
(see derivation in Appendix A), which is equivalent to the corresponding expression (11) of \cite{Mark71}. The subscript in the l.h.s. reveals the type of the bypass: `${\frown}$' stands for bypass from above. With help of the integral representation for the modified Bessel function \citep{GR15}
\be
I_1 (z)\,\exp(-z)=\pi^{-1}\int_0^{\pi}\exp[-z\,(1-\cos\theta)]\,\cos\theta\,\d\theta
\label{eq:bessel}
\ee
an alternative expression for ${\cal F}_{\frown}^{\rm RES}$ is:
\be
{\cal F}_{\frown}^{\rm RES}(y,u_L)=\frac{\ell}{r\epsilon x}\,F^{\rm RES}_{\frown},
\ee
where
\begin{multline}
F^{\rm RES}_{\frown}=
\frac{1}{\pi}\int\limits_0^{\pi} e^{-x\,(1-\cos\theta)}\cos\theta \,\d\theta\\
\times
\Bigl[e^{-Z^2/2}\Bigl(\int\limits_0^Z e^{s^2/2} \d s-i\,\sqrt{\frac{\pi}{2}}\Bigr)\Bigr],
\label{eq:F_res_frown}
\end{multline}
and  $Z \equiv u_L+i\,y\,(1-\cos\theta)$, $x \equiv y^2$.

The nonresonant contribution ${\cal F}^{\rm NR}$ can be simplified by setting $\nu=-1$ and summing up the series:
\begin{multline}
    {\cal F}^{\rm NR}= \frac{e^{-x}}{x}
\sum\limits_{n\ne -1}\frac{n\,I_n(x)}{n-\nu}\approx \frac{1}{x}-\frac{1}{x^2}\\
+I_0(x)\,e^{-x}\,
\left(-\frac{1}{x}+\frac{1}{x^2}\right)+I_1(x)\,e^{-x}\,
\left(-\frac{3}{2\,x}+
\frac{2}{x^2}\right)+\\
+I_2(x)\,e^{-x}\,
\left(\frac{1}{x}+\frac{2}{x^2}\right)+I_3(x)\,e^{-x}\,\left(\frac{1}{2x}\right).
\label{eq:NR}
\end{multline}
After substitution, one obtains the dispersion relation in the form
\begin{multline}
y^2= y_L^2 \,(2 y s_y)\,\frac{1}{\pi}\int\limits_0^{\pi} e^{-x\,(1-\cos\theta)}\cos\theta \,\d\theta\\
\times\Bigl[e^{-Z^2/2}\Bigl(\int\limits_0^Z e^{s^2/2} \d s-i\,\sqrt{\frac{\pi}{2}}\Bigr)\Bigr]
+s_y\,\frac{y}{y_T}\, \bigl[x\, {\cal F}^{\rm NR}(x)\bigr],
\label{eq:DEMark}
\end{multline}
where  $s_y={\rm sgn}[{\rm Re}(y)]=s_k$,  $y_T=k_T\,(r\epsilon)$, $y_L^2=k_L^2\,(r\epsilon)^2$,
$k_L^2= \pi G\sigma_0\varkappa^2 \ell/{c^4}$. Note that $y_L$ is a large parameter,
$$
y_L^2\sim \frac{\ell}{ c/\varkappa}\cdot\frac{1}{Q}\sim\frac{1}{\epsilon\,Q}\gg 1,
$$
provided the Toomre stability parameter $Q=c\varkappa/(3.36G\sigma_0)$ \citep{Toomre64} is of order 1. This means that in the resonance region a typical wavenumber $k$ is large not only compared to the inverse size of the inhomogeneity $\sim r^{-1}$, but also to the typical inverse size of the epicycles $a^{-1}$, i.e. the dimensionless wavenumber $ y\equiv k(c/\varkappa)=k\,(r\epsilon)$ is large, $|y|\gg 1$. This in turn allows to simplify the r.h.s. of (\ref{eq:DEMark}), and also omit the nonresonant contribution in our approximation. Then (\ref{eq:DEMark}) gives (see details of derivation in Appendix B):
\be
y^2=y_L^2\,e^{-u_L^2/2}\,\Biggl\{- i\,\Bigl[1+(s_y-1)\,e^{-2iu_{L}y}\Bigr] +
\sqrt{\frac{2}{\pi}}\int\limits_0^{u_{\!L}}\d\,s\,e^{s^2/2}\Biggr\}.
\label{eq:DEMark_asymp}
\ee

The last equation must be equivalent to equation (11) of \cite{Mark71}, since both were obtained under the same assumptions. However, comparison shows that they coincide only for the case of trailing spirals, $s_y=+1$. In this case from (\ref{eq:DEMark_asymp}) we have:
\be
y^2=y_L^2\,e^{-u_L^2/2}\,\Bigl(-i+\sqrt{\frac{2}{\pi}}\int\limits_0^{u_{L}} \d s\,e^{s^2/2}\Bigr).
\label{eq:DEMark_asymp_p}
\ee
Solution of this equation has a negative imaginary part, $y_i<0$, corresponding to trailing wave decay towards the disc centre inside the resonance region, in full agreement with \cite{Mark71, Mark74}. In particular, for $u_L\gg 1$ we find:
\be
y_r=\Bigl(\frac{2}{\pi}\Bigr)^{1/4}\,\frac{y_L}{u_L^{1/2}},
\label{eq:DEMark_asymp_p_real}
\ee
\be
y_i=-{\frac{1}{2}}\,y_L\,\Bigl(\frac{\pi}{2}\Bigr)^{1/4} u_L^{1/2} e^{-u_L^2/2}.
\label{eq:DEMark_asymp_p_im}
\ee

In the case of leading spirals, $s_y=-1 $, from (\ref{eq:DEMark_asymp}) one obtains:
\be
y^2=y_L^2\,e^{-u_L^2/2}\,\Bigl[-i\,\Bigl(1-2\,e^{-2iu_{L}y}\Bigr) +
\sqrt{\frac{2}{\pi}}\int\limits_0^{u_{L}} \d s\,e^{s^2/2}\Bigr].
\label {eq:DEMark_asymp_m}
\ee
This equation occurs substantially different from the eq. (11) of {Mark's work (1971)} (which has the same form for leading and trailing spirals, namely, Eq. (\ref{eq:DEMark_asymp_p}) in our notation). Thus Mark's equation gives the complex wavenumbers $ k=k_r+ik_i $ for leading and trailing spirals that differ in sign only. Meanwhile, from comparison of (\ref{eq:DEMark_asymp_m}) and (\ref{eq:DEMark_asymp_p}) we infer that the difference is more significant.\footnote{The Mark's form of the equation for leading spirals is obtained using inaccurate simplification of the eq. (7) of \cite{Mark71} or eq. (40) in \cite{Mark74}, equivalent to our equation (\ref{eq: RF_Mark_a}): neglecting of the second term in r.h.s. of the asymptotic expansion ($|z|\gg 1$) of the function $I_1(z)\,e^{-z}$:
$$
I_1(z)\,e^{-z}\sim\frac{1}{\sqrt{2\pi z}}\,\bigl[1+{\cal O}(1/z)\bigr] +
\frac{\exp[-2\,z\pm\frac{3}{2}\,\pi\,i]}{\sqrt{2\pi z}}\,\bigl[1+{\cal O}(1/z)\bigr].
$$
\citep[see formula (8.452.5) ​​in][]{GR15}. This term is significant in the case of leading spirals and gives a comparable contribution.} In particular, at the periphery of the resonance region, $u_L\gg1$, where $y_i$ is exponentially small, we find by perturbation theory
\be
y_r=-\Bigl(\frac{2}{\pi}\Bigr)^{1/4}\,\frac{y_L}{u_L^{1/2}},
\label{eq:DEMark_asymp_m_real}
\ee
\be
y_i={\frac{1}{2}}\,y_L\,\Bigl(\frac{\pi}{2}\Bigr)^{1/4}u_L^{1/2}e^{-u_L^2/2}\,\bigl[1-2\cos(2u_L y_r)\bigr].
\label{eq:DEMark_asymp_m_im}
\ee
Interestingly, {\it averaging} of expression (\ref{eq:DEMark_asymp_m_im}) over the oscillations gives Mark's expressions for leading spirals, meaning that Mark's conclusion concerning the leading waves decay outwards the resonant circle is generally valid.

\begin{figure*}
 \begin{center}
 \includegraphics[width=170mm]{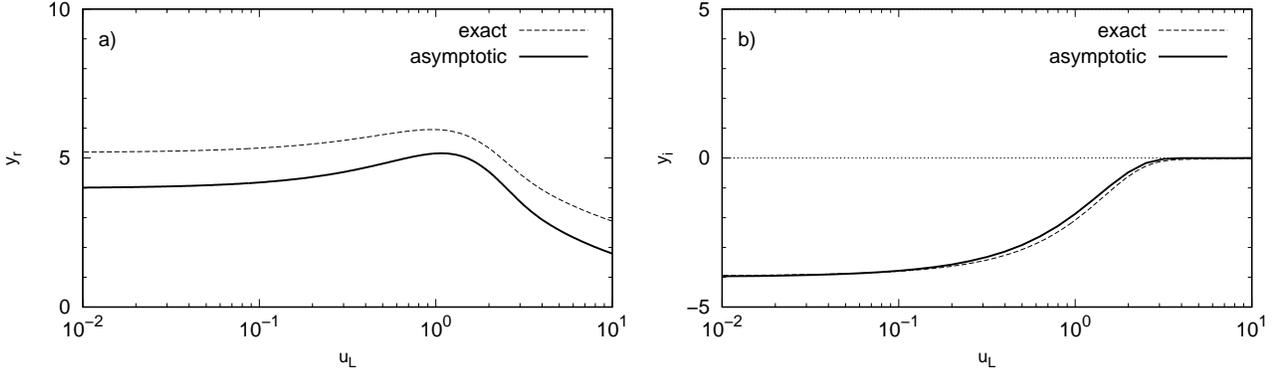}
 \caption{\protect\footnotesize --  The real (a) and imaginary (b) parts of the complex trailing solutions $y=y_r+i\,y_i$ for $y_L=5.64$ calculated from the exact equation (\ref{eq:DEMark}) taking into account the nonresonant contribution (thin line), and from the asymptotic equation (\ref{eq:DEMark_asymp_p}) (thick line).}
 \label{Fig.1}
 \end{center}
\end{figure*}

Fig.\,\ref{Fig.1} shows the real and imaginary parts of the complex trailing solutions of the exact dispersion relation (\ref{eq:DEMark}) and obtained from asymptotic expression (\ref{eq:DEMark_asymp_p}). The model parameters are adopted from \cite{Mark74}: $ y_L = 5.64$, $\Omega_{\rm p}=13.5$ km\! s$^{-1}$\! kps$^{-1}$, $R_L=3.2$ kps, $\ell=15.4$ kps, $a_L\equiv(c/\varkappa)_L = 0.64$ kps, $k_L=8.8$ kpc $^{-1}$. The solutions are completely consistent with those obtained by Mark.

It is worth noting that approaching the resonance, the imaginary part of the wavenumber increases in the absolute value and becomes comparable to its real part, $|k_i(r)|\sim |k_r (r)|$, i.e. at $r \approx R_L$ the assumption $|k_i|\ll|k_r|$ fails. The latter is used, in particular, in derivation of the leading order WKB relation connecting the amplitudes of the perturbed potential and surface density:
$$
 V=-\frac{2\pi G S}{|k_r|}\equiv -\frac{2\pi G S}{k_r}\,s_k.
$$
Strictly speaking, the dispersion relation and the solutions are valid only in the range of their applicability, $|y_i|\ll|y_r|$. In essence, we restricted to use asymptotic expressions (\ref{eq:DEMark_asymp_p_real}) and (\ref{eq:DEMark_asymp_p_im}) for trailing waves, and (\ref{eq:DEMark_asymp_m_real}) and (\ref{eq:DEMark_asymp_m_im}) for the leading waves.

The issue of formal inapplicability of the equations near to the resonance circle (closer than the size of the epicycle) was already mentioned by Mark, when he tried to reconcile his solution for trailing waves with the fundamental paper by \citet[][hereafter LBK]{LBK72}. The paper contains an overall  (over the whole disc) angular momentum exchange rate $\d\langle L \rangle/\d t$ between the {\it stationary} wave and stars at resonances obtained without using the WKB and epicyclic approximations. Mark's and LBK solutions agree only if the asymptotic expression (\ref{eq:DEMark_asymp_p_im}) for $k_i(r)$ is continued down to $r=R_L$, or $u_L(r)=0$ (see Sec. IV and Appendix B in \cite{Mark74}). Below we will discuss it in more detail.

For the leading waves, such a continuation of the solution of our equation (\ref{eq:DEMark_asymp_m}) to a region sufficiently close to $r=R_L$ gives not only comparable, but also arbitrary large $|y_r|$ and $y_i$. Therefore, for leading spirals the inapplicability is even more pronounced. \cite{Mark71, Mark74} did not realised this difficulty because his dispersion relation for leading spirals was inaccurate.\footnote{At first glance such an asymmetry of expressions for $s_k=+1$ and $s_k=-1$ is puzzling. The reason is the direction of the bypass,  which is determined by the sign of the derivative $(\d\nu/\d R)_{R=R_L}$. In our problem, i.e. real galaxies, $(\d\nu/\d R)_{R=R_L}>0$ suggests the bypass of the pole in the complex plane from above. In the case $ (\d\nu/\d R)_{R=R_L}<0$, we would get the bypass from below, and the nonmonotonic behaviour of $y_i$ would be in the trailing rather than in the leading spirals.}

\section{Dispersion relation for the modified bypass rule}

Based on the analogy with critical layers in hydrodynamics discussed in Section 2, here we investigate the consequences of the hypothesis that the weak nonlinearity in the resonance region leads to a modification of the standard bypass rule. Reviews of the relevant papers can be found in \citet{Maslowe86} and \citet{CS96a, CS96b}. This analogy dictates a replacement of the Landau\,-\,Lin bypass rule for integration to calculation in the principal value sense.

\smallskip
The modified dispersion relation is derived in Appendix A. Since the principal value integral is equal to a half-sum of integrals with bypass from above and from below, we find instead of ${\cal F}^{\rm RES}$:
\be
{\cal F}^{\rm RES}_P(y,u_L)= \frac{\,\ell}{r\,\epsilon\,x}\, F_P^{\rm RES},\ \
F_P^{\rm RES}= \frac{F_{\frown}^{\rm RES}+F_{\smile}^{\rm RES}}{2},
\label{eq:F_half-sum}
\ee
where $F_{\frown}^{\rm RES}$ is defined by (\ref{eq:F_res_frown}), index `\,$\smile$' denotes the bypass from below, `$P$' -- calculation of the integral in the principal value sense, and
\be
 F_{\smile}^{\rm RES}=-i\int\limits_{+\infty}^0 I_1(x-\mu\,y)\,\,e^{-(x-\mu\,y)}\,e^{-\frac{1}{2}\,\mu^2-i\,u_L\,\mu}\,\d\mu.
\ee
Using integral representation (\ref{eq:bessel}), we obtain for $F_{\smile}^{\rm RES}$:
\begin{multline}
F_{\smile}(y,u_L)=
\frac 1{\pi}\int\limits_0^{\pi} e^{-y^2\,(1-\cos\theta)}\cos\theta \,\d\theta\\
\times \Bigl[e^{-Z^2/2}\Bigl(\int\limits_0^Z e^{s^2/2} \d s+i\,\sqrt{\frac{\pi}{2}}\Bigr)\Bigr],
\label{F_res_smile}
\end{multline}
so that the half-sum is
\begin{multline}
{\cal F}^{\rm RES}_P=\frac{\ell}{r\,\epsilon x}\cdot\frac{F_{\frown}^{\rm RES}+F_{\smile}^{\rm RES}}{2}\\
=\frac{1}{\pi}\int\limits_0^{\pi} e^{-y^2\,(1-\cos\theta)}\cos\theta \,\d\theta
\times \Bigl(e^{-Z^2/2}\int\limits_0^Z e^{s^2/2} \d s\Bigr).
\label{eq:half-sum}
\end{multline}

Comparison of the derived reduction factor (\ref{eq:half-sum}) with its counterpart (\ref{eq:F_res_frown}) for the Landau\,-\,Lin bypass rule shows that the second term in square brackets in r.h.s. of (\ref{eq:F_res_frown}), $\sim -i\,(\pi/2)^{1/2}\,e^{-Z^2/2}$, is due to the contribution of the semi-residue associated with the bypass of the pole. Importantly, the removal of this contribution does not make the reduction factor $ {\cal F}_{\rm P}^{\rm RES}$ real for real $y$ (i.e. for real $k$), because parameter $Z=u_L+i\,y\,(1-\cos\theta)$ remains complex even for real $y$. Thus, the modified dispersion relation will not result in vanishing of the imaginary part of the wavenumber $k_i$, which seemingly means presence of dissipative effects associated with the resonance exchange of the angular momentum and energy between the wave and the stars.\footnote{In the electron plasma, absence of damping obtained by \citet{Vlasov1945} is solely due to integration in the principal value sense \citep{Landau46}. On the other hand, small but finite amplitude of the wave (i.e., a weak nonlinearity) leads to appearance of trapped electrons and formation of a plateau in the distribution function.  This fact justifies Vlasov's approach and undamped oscillations obtained in his work (this remark was made by editors of a special issue of `Uspekhi Fizicheskih Nauk' (1967) devoted to the famous Landau paper.)} Let's consider this situation in more detail.

Substitution of (\ref{eq:half-sum}) into r.h.s. of dispersion relation (\ref{eq:stand1}) (the nonresonant part is omitted) gives instead of (\ref{eq:DEMark}):
\begin{multline}
y^2= y_L^2 \,(2 y s_y)\,\frac{1}{\pi}\int\limits_0^{\pi} e^{-x\,(1-\cos\theta)}\cos\theta \,\d\theta\\
\times\Bigl(e^{-Z^2/2}\int\limits_0^Z e^{s^2/2} \d s\Bigr).
\label{eq:DE_modif}
\end{multline}
In the limit ${\rm Re}(x)\gg 1$, it is simplified to (see Appendix B):
\be
y^2/y_L^2=e^{-u_L^2/2}\,\Bigl(-i\,s_y\,e^{-2\,iu_{L}y} +
\sqrt{\frac{2}{\pi}}\int\limits_0^{u_{L}} \d s\,e^{s^2/2}\Bigr).
\label{eq:ep1ps2}
\ee
It is now obvious that roots of the dispersion relation in case of the modified bypass rule cannot be purely real. Outside the resonance region from (\ref{eq:ep1ps2}) one finds:
\be
|y_r|=y_L\,\Bigl(\frac{2}{\pi}\Bigr)^{1/4}\,u_L^{-1/2},
\label{eq:modif y_r_asymp}
\ee
\be
y_i=-\frac12\,y_L\,u_L^{1/2}\,\Bigl(\frac{\pi}{2}\Bigr)^{1/4}\cos(2 u_L y_r)\,e^{-u_L^2/2}.
\label{eq:modif y_i_asymp}
\ee
\begin{figure*}
 \begin{center}
 \includegraphics[width=170mm]{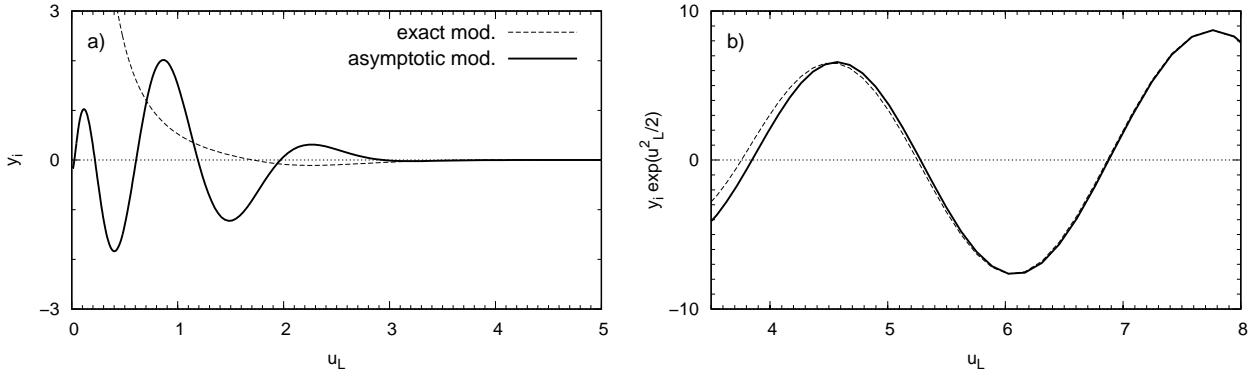}
 \caption{\protect\footnotesize
  Solution of the eq. (\ref{eq:modif y_i_asymp}) for the modified bypass rule and $y_L=5.64$ (thick lines): (a) imaginary part $y_i(u_L)$ and (b) magnification $y_i(u_L) \exp(u_L^2/2)$. Thin lines show $y_i$ obtained from the solution of (\ref{eq:ep1ps2}).}
 \label{Fig.2}
 \end{center}
\end{figure*}

Fig.\,\ref{Fig.2} shows that $y_i$ oscillate toward the resonance, so that regions of decay  $k_i<0$ alternate with the regions of amplitude growth $k_i>0$. Physically this means that regions in which the wave transfers the angular momentum to the stars alternate to the regions where the stars transfer the angular momentum to the wave. This holds both for leading and trailing waves. Assuming that the asymptotic expression  (\ref{eq:modif y_i_asymp}) is valid  up to the resonance circle, one obtains almost complete compensation for the dissipative effects. For example, for $y_L = 5.64$ the integral $\int_0^{\infty} y_i(u_L)\,\d u_L\sim 10^{-2} $, and for $y_L=10$ this integral $\sim 10^{-4}$. In other words, the overall amplification coefficient $\Gamma$ is close to unity,
\be
\Gamma = \exp \Bigl[\int_{R_L}^{R_L+\Delta r} k_i(r)\,\d r\Bigr] \approx
\exp\Bigl[\int_{0}^{\infty} y_i(u_L)\,\d u_L\Bigr]\approx 1,
\label{eq:Gamma}
\ee
i.e. no wave amplification or attenuation. Here we assume $\Delta r\gg (c/\varkappa)$.

Unfortunately, numerical solution of equation (\ref{eq:DE_modif}) and the analytical analysis of nearly equivalent approximate equation (\ref{eq:ep1ps2}) show an unbounded growth of $y_i$ as one approaches $u_L = 0$, meaning the unlimited amplification of the wave in the direction of the resonance. Formally, this is due to the term $\exp(-2iu_Ly)$ in the r.h.s. of equation (\ref{eq:ep1ps2}). We have already seen this behaviour in Sec. III analysing the leading spirals, and obviously for the same reason: violation of the assumptions made near the resonance circle. Therefore, our conclusion concerning $\Gamma\approx 1$ based on the asymptotic equation (\ref{eq:modif y_i_asymp}) may seem unconvincing. Nevertheless, there is another supporting evidence considered in the next section.

\section{The overall effect of the resonance region}

\cite{Mark74} generalised the wave action balance equation, derived earlier by \cite{Shu70} \cite[see also Sec. 6.2.6 in][]{BT2008} by adding the wave action source associated with the transfer of wave action from the stars to the wave:
\be
 \frac{\p{\cal U}}{\p t}+\frac{1}{r}\,\frac{\p}{\p r}\,(r{\cal B}) = {\cal S}.
\ee
Here ${\cal U}(r,t)$ is the wave action density, ${\cal B}(r,t)$ and ${\cal S}(r,t)$ are the flux and source of the wave action. We recall that the wave action is related to the angular moment density ${\cal L}$ and the energy density ${\cal E}$ of the wave as ${\cal L}=m\,{\cal U}$ and ${\cal E}=\omega\,{\cal U}$, respectively. The transfer rate of the angular momentum between the resonance stars and the wave is:\footnote {In case of a stationary wave, dissipation on the resonance suggests a source  ${\cal S}_{\rm ext} $ somewhere outside the resonance that compensate the absorption.}
\be
\frac{\d\langle L\rangle}{\d t} = 2\pi\int_0^{\infty}\frac{\d{\cal L}(r,t)}{\d t}\,r\,\d r =
2\pi m \int_0^{\infty} {\cal S}(r,t)\,r\,\d r.
\label {eq:dLdt}
\ee
In fact, the integration over $r$ in (\ref{eq:dLdt}) is carried out only in the neighbourhood of the resonance, where the source $ {\cal S} $ is nonzero. The local source of the angular action ${\cal S} $ and of the angular momentum $m\,{\cal S}$ is related to $k_i(r)$ as follows:
\be
m\,{\cal S}(r)=\frac{m\,s_k}{2\pi G}\,k_i(r)\,|V(R_L)|^2,
\label {eq:mS}
\ee
where $V$ is the amplitude of the wave potential \citep[see eqs. (65), (69), and (72) in][]{Mark74}.

On the other hand, LBK calculated the overall (integrated over the whole resonance region) transfer rate of the angular momentum to the wave from the resonance stars. Significantly, they didn't use WKB approximation, but only assumed the adiabatic `turn-on' of the wave in the distant past ($t\to - \infty$). This formally corresponds to multiplication of the wave amplitude to additional factor $\e^{\delta\cdot t}$ with small $\delta>0$. The procedure is equivalent to the causality principle leading to the standard bypass rule in a purely linear problem. Therefore, eqs. (\ref{eq:dLdt}) and (\ref{eq:mS}) for the integral change of the wave angular momentum due to the resonance stars, obtained in the WKB approximation in the present works and \citet{Mark74}, should coincide with minus change of angular momentum of resonance stars $d \langle L\rangle/dt$ obtained by LBK.

The angular momentum change of stars near ILR for the bisymmetric spiral $m=2$ is (LBK):
\begin{multline}
\Bigl(\frac{\d\langle L\rangle}{\d t}\Bigr)_{n=-1}=\frac{1}{4\pi}\int\int\Bigl(-\frac{\p F}{\p I_1}
+2 \,\frac {\p F}{\p I_2}\Bigr)\,|\phi_{-1}(I_1,I_2)|^2\\
\times\delta(\omega+\Omega_1-2\Omega_2)\,\d I_1\, \d I_2.
\label{eq:LBK}
\end{multline}
where $I_1$ and $I_2\equiv L = rv_{\varphi}$ are the radial and azimuthal actions
of the unperturbed orbit (in terms of action\,-\,angle variables), $\Omega_1$ and $\Omega_2$ are the corresponding frequencies, $F(I_1,I_2)$ is the equilibrium DF, $\phi_n=\int_0^{2\pi}\int_0^{2\pi}
 \d w_1\,\d w_2\,V [r (w_1)\,\exp\bigl[\int^{r(w_1)}k(r')\,dr'\bigr]\,\exp [i\,(m\varphi-m w_2 -nw_1)]$, and $(w_1,w_2)$ are the angular variables conjugate to $(I_1,I_2)$.

The delta function $\delta(\omega+\Omega_1-2\Omega_2)$ in (\ref{eq:LBK}) occurs from a semi-residue of the pole $(\omega+\Omega_1-2\Omega_2)^{-1}$ originated from the limit $\delta\to+0$ and the bypass of the singularity. In the epicyclic approximation, the pole is transformed into $\{\varkappa\,[\nu(R)+1]\}^{-1}$, and the delta function is transformed into $\delta(R-R_L)/(\varkappa\,|d\nu/dR|_{R=R_L})$. Assuming the perturbation of the potential in the form (\ref{eq:WKB}), and DF (\ref{eq:DF}), one can obtain from (\ref{eq:LBK}) \citep[see details in Appendix B of][]{Mark74}:
$$
\Bigl(\frac{\d\langle L \rangle}{\d t}\Bigr)_{-1}=2\pi\,m\int_{R_L}^{\infty}{\cal S}_{\rm LBK}(r)\,r\, \d r,
$$
and
\be
 {\cal S}_{\rm LBK}(r)=\frac{\ell\,\sigma_0(R_L)}{4\varkappa^2(\epsilon_L R_L)^4}
\,\frac{|V(R_L)|^2}{|k(r)|}\,\exp\Bigl[-\frac{1}{2}\,\Bigl(\frac{r-R_L}{R_L\epsilon_L}\Bigr)^2\Bigr].
\label{eq:S_LBK}
\ee

It can be checked explicitly with help of the dispersion relation (\ref{eq:ep1ps2}) that the wave overall angular momentum loss rate (\ref{eq:dLdt}) and (\ref{eq:mS}) corresponds to the integral angular momentum acquired by the resonance stars (\ref{eq:S_LBK}). This is the case, however, only if one extends the asymptotic expressions (\ref{eq:DEMark_asymp_p}) and (\ref{eq:DEMark_asymp_p_real})  obtained for the Landau-Lin bypass rule \citep[and equivalent to the corresponding expressions by][]{Mark71, Mark74} down to the resonance circle $r=R_L$. This quantitative agreement of two different approaches, one of which (LBK) does not use the WKB approximation, supports the conclusion that the asymptotic WKB expressions for $k_i(r)$ correctly describes dissipative processes in the entire region.

Based on this agreement, we suppose that the asymptotic expressions (\ref{eq:modif y_r_asymp}) and (\ref{eq:modif y_i_asymp}) for $k_r$ and $k_i$  are adequate to describe the overall wave interaction for the modified bypass rule as well. The estimates (\ref{eq:Gamma}) shows neither amplification nor decay of the wave. The same follows from the balance equation (\ref{eq:dLdt}), if one uses (\ref{eq:modif y_i_asymp}) in the expression for ${\cal S}$:
  \begin{multline}
\frac{d\langle L\rangle}{dt} =
2\pi m\int_0^{\infty}{\cal S}(r,t)\,r\,\d r \\ \approx s_k\,\frac{2\,R_L\,|V(R_L)|^2}{G}
\int_{R_L}^{R_L+\Delta r}k_i(r)\,\d r\approx 0.
\label{eq:int_ki}
\end{multline}

Concluding, the modification of the bypass rule does not prevent the local angular momentum exchange near the resonance, $k_i(r)\ne 0$, but the overall exchange is zero. This is in agreement with LBK approach, if the bypass rule in their theory is modified accordingly.

\section{Discussion}

1. We consider a problem of the spiral wave interaction with stars near ILR in galactic discs, in the leading orders of the WKB and epicyclic approximations. In the purely linear formulation, when nonlinear effects do not play any role even near the resonance, the problem was considered earlier by \citet{Mark71, Mark74}. Using the action\,-\,angle technique, we rederive his result for the trailing spirals, and improved treatment of leading waves. It is turned out, however, that this improvement doesn't make a significant change to Mark's physical conclusion about the fate of leading spiral density waves.
\smallskip

2. Using an analogy with critical levels in hydrodynamics, we investigate the spiral wave interaction in the case of weak nonlinearity, assuming that the standard Landau\,-\,Lin bypass rule of the pole for the purely linear problem is changed to the calculation of corresponding integral in the principal value sense. A similar modification is also known in the theory of electron plasma oscillations, where trapped particles lead to the formation of a plateau on the distribution function and the change of the bypass rule.

The principal-value approach to the resonance integral is heuristic in nature and does not necessarily have the physical back-reaction mechanism, so the full reflection of trailing waves to leading ones is an open question. Rigorous justification of the bypass rule modification requires a rather laborious investigation of particle dynamics and the structure of DF in the gravitational field of the spiral wave near ILR. Such an attempt was made for the first time by \cite{Contopoulos70b} (mainly numerically) and some years later (mainly analytically) by \cite{Vandervoort73, Vandervoort75} and \cite{VM75}. However, it is unclear so far how to apply their results to our study. It is also worth mentions the work by \cite{GT81}, who considered perturbations of particle trajectories in Uranus rings caused by a satellite near ILR.

We show that the modified bypass rule does not prevent the angular momentum exchange between the wave and stars at all radii $r$ near the resonance, in contrast to similar problems in hydrodynamics. For example, \cite{Stew78}, \cite{WW78}, and especially the work by \cite{KM85} with a picturesque title ``Do Rossby-wave critical layers, reflect, or over-reflect?\,'' analyse interaction of Rossby waves with the critical layer and show that nonlinearity in neighbourhood of the resonance turns off the wave absorption taking place in a purely linear problem \citep[as it was shown by][]{Dick70}, and result in full reflection. In plasma physics, this bypass rule also leads to ``switching off'' the Landau damping.

Nevertheless, the absorption on ILR disappears integrally over the resonance region: radial intervals of the wave amplification alternate with the intervals of the wave decay. Mutual compensation of amplification and decay occurs, so that the overall effect is zero. Moreover, this result is valid both for leading and trailing spirals.
The resonance in this case does not remove the problem posed by the antispiral theorem \citep{LBO67}, since the modified bypass rule actually eliminates the temporal irreversibility of dynamical equations due to the resonance taking place for the standard bypass rule.

The initial motivation for this work was the problem of the feedback in over\,-\,reflection theories of galactic spiral structure. We show that a weak nonlinearity can prevent wave absorption and produce the needed feedback loop of the transformation of trailing waves to leading. Following \citet{Mark71, Mark74}, we stress here on ILR, although the same effect of decay removal takes place on OLR.

Another motivation was theoretical explanation of the well-known fact that bars in numerical experiments form with pattern speeds that avoid ILR, either because of the fast rotation, or due to the presence of a so-called $Q$-barrier isolating the resonance \citep[e.g.][]{CE93, PBJ2016}. Numerical bars can be traced back evolutionary to the most unstable very open trailing spiral modes \citep{Pol2013}, so Mark theory cannot be used here directly since it requires extension from tightly wound to open spirals. Besides, calculations in fluid discs \citep{Pol2018} also show absence of global modes with ILR, although fluid waves are likely insensitive to this resonance. This issue will be addressed elsewhere.

\section*{Acknowledgements}

We thank the referee, Frank Shu, for suggestions on improving the original version of the paper. This work was supported by the Sonderforschungsbereich SFB 881 ``The Milky Way System'' (subproject A6) of the German Research Foundation (DFG), and by the Volkswagen Foundation under the Trilateral Partnerships grant No. 90411. The authors acknowledge financial support by the Russian Basic Research Foundation, grant 16-02-00649, and by Department of Physical Sciences of RAS, subprogram `Interstellar and intergalactic media: active and elongated objects'.  The work also was partially performed with budgetary funding of Basic Research program II.16 (Ilia Shukhman).

\appendix

\section{The surface density response and the reduction factor near ILR. Epicyclic approximation and action\,-\,angle technique}

In the epicyclic approximation, the action variables in axisymmetric potential $\Phi_0(r)$ are
$$
I_1\equiv I=\case{1}{2}\,\varkappa(R)\,a^2, \ \ \ I_2 \equiv L = \Omega(R)\,R^2,
$$
where
$$
\Omega(R)\equiv\Omega_2 =\frac{1}{R}\,\frac{d\Phi_0(R)}{dR},$$ $$\varkappa(R)\equiv \Omega_1=\sqrt{4\Omega^2+2R\Omega(R)\Omega'(R)}
$$
are the angular velocity of the guiding centre of the epicycle at radius $R$ and the epicyclic frequency, respectively; $a$ is the size of the epicycle supposed to be small, $a\ll R$. The angular variables $w_1$ and $w_2$ conjugated to action variables $I_1$ and $I_2$ are related to the current radial and azimuthal coordinates of the star $r$ and $\varphi$ as follows:
\begin{equation}
r=R-a\cos w_1,\ \ \ \varphi=w_2 +{\cal O}(a/R),
\label{eq:r-phi}
\end{equation}
and obeying the equations
\begin{equation}
\frac{d w_1}{dt}=\Omega_1(R,a)\approx\varkappa(R),\ \ \ \ \frac{d w_2}{dt}=\Omega_2(R,a)\approx \Omega(R).
\label{eq:dw_12}
\end{equation}

In the leading order of the perturbation theory adopted here, the small parameter $a/R$ and the WKB parameter $(kR)^{-1}\sim (kr)^{-1}$ are considered to be of the same order, ${\cal O}(a/R)\sim (kr)^{-1}\ll 1$. The contribution of the order ${\cal O}(a/R)$ in the expression (\ref{eq:r-phi}), which connects $w_2$ and coordinates $(\varphi,r)$, can be ignored. One can use pairs $(I,R)$ or $(a,R)$ instead of the action variables  $I$ and $L$ if more appropriate. Besides, in the leading order of the epicyclic approximation the differences $\Omega_1(R, a)-\varkappa (R)$ and $\Omega_2(R, a)-\Omega(R)$ are negligible, since they are of the order of ${\cal O}(a^2/R^2)$ \citep{Shu70,Mark76,B14}.

The epicyclic  energy ${\cal E}\equiv E-E_{\rm circ }(R)$, where
$$
E={\case{1}{2}}\,(v_r^2+v_{\varphi}^2)+\Phi_0(r)={\case{1}{2}}\,
\Bigl(v_r^2+\frac{L^2}{r^2}\Bigr)+\Phi_0(r),
$$
$$
E_{\rm circ}={\case{1}{2}}\, [\Omega(R)\,R]^2+\Phi_0(R)
$$
can be written as
\begin{equation}
{\cal E}={\case{1}{2}}\,(v_r^2+\tilde{v}_{\varphi}^2)=\varkappa\,(R)\,I={\case{1}{2}}\, [\varkappa\,(R)\,a]^2
\label{eq:energy}
\end{equation}
where
\begin{equation}
\tilde{v}_{\varphi}=\frac{2\Omega(R)}{\varkappa\,(R)}\,[v_{\varphi}-\Omega(r)\,r)]=
-\varkappa\,(R)\,(r-R).
\label{eq:v_phi}
 \end{equation}
The equilibrium DF in the epicyclic approximation is supposed to be of the Schwarzschild type:
\begin{equation}
F(r,v_r,v_{\varphi})=P(R)\,\exp\Bigl[-\frac{\cal E}{c^2(R)}\Bigr],
\label{eq:DF_A}
\end{equation}
where
\begin{equation}
P(R)=\frac{2\Omega(R)}{\varkappa(R)}\,\frac{\sigma_0(R)}{2\pi\,c^2(R)},
 \end{equation}
$\sigma_0(r)$ is the equilibrium surface density, and $c(r)$ is the radial velocity dispersion.
Using (\ref{eq:energy}) and (\ref{eq:v_phi}), the DF can be also written as \citep[e.g.][]{Contopoulos71}:
\begin{equation}
F=P(R)\,\exp\Bigl[-\frac{v_r^2+\varkappa^2(R)\,(r-R)^2}{2\,c^2(R)}\Bigr].
\label{eq:DF_Cont}
 \end{equation}
In the variables
$$
\xi \equiv \frac{v_r}{R\,\varkappa(R)}, \ \ \eta \equiv 1-\frac{r}{R},
 $$
DF takes the form:
\begin{equation}
F=P(R)\,\exp\Bigl[-\frac{\xi^2+\eta^2}{2\,(c/\varkappa\,R)^2}\Bigr].
 \end{equation}
The element of the volume in the velocity space
is
\begin{equation}
\d\Gamma\equiv \d v_r\,\d v_{\varphi}=\frac{\varkappa^3\,R^2}{2\Omega}
\,\d\xi\,\d\eta = \frac{\varkappa^3}{2\Omega}\,a\,\d a\,\d w_1,
\label{eq:dGamma}
 \end{equation}
where
$\xi=({a}/{R})\,\sin w_1,\ \ \eta= ({a}/{R})\,\cos w_1.
$
Here one can use $r$ and $R$ interchangeably.

\smallskip
The linearised collisionless Boltzmann equation for the perturbed DF $\delta f(I_1,I_2,w_1,w_2)$ in the action\,-\,angle variables is
\be
\frac{\p\,\delta\!f}{\p t}+\Omega_1\,\frac{\p\,\delta\!f}{\p w_1}+\Omega_2\,\frac{\p\,\delta\!f}{\p w_2}=
\frac{\p\,\delta\Phi}{\p w_1}\,\frac{\p F}{\p I_1}+\frac{\p\,\delta\Phi}{\p w_2}\,\frac{\p F}{\p I_2}.
\label{eq:BE}
\ee
For the bisymmetric ($m=2$) perturbations the perturbed potential
\be
\delta\Phi(r,\varphi,t)= V(r)\,\exp\Bigl\{i\,\Bigl[\int\limits^r k(r')\,\d r'+2\varphi-2\Omega_{\rm p} t\Bigr]\Bigr\}
 \label{eq:delta_Phi}
\ee
can be written as:
\be
\delta\Phi=e^{2i\,(w_2-\Omega_{\rm p} t)}
\sum\limits_{l=-\infty}^{\infty} \phi_l(I_1,I_2)\,e^{i l w_1},
 \label{eq:delta_Phi_1}
\ee
and from (\ref{eq:BE}) one can have:
$$
\delta f=-\,e^{2i\,(w_2-\Omega_{\rm p} t)}
\sum\limits_{l=-\infty}^{\infty}
\frac{l\,\dfrac{\p F}{\p I_1}+2\,\dfrac{\p F}{\p I_2}}
{2(\Omega_{\rm p}-\Omega_2)-l\Omega_1}\, \phi_l(I_1,I_2)\,e^{i l w_1}.
$$
Changing the variables $(I_1,I_2,w_1,\varphi)$ to $(R,a,w_1,\varphi)$, taking into account that
$\p F/\p I_1\gg \p F/\p I_2$  and denoting
\be
\nu(R)=\frac{2 [\Omega_{\rm p}-\Omega(R)]}{\varkappa(R)},
\label{eq:nu}
\ee
we obtain the perturbed DF
\be
\delta f=-\frac{e^{2i(\varphi-\Omega_{\rm p}t)}}{\varkappa^2}
\sum\limits_{l=-\infty}^{\infty}
\frac{(l/a)\,{\p F}/{\p a}}
{\nu(R)- l}\, \phi_l(R,a)\,e^{i l w_1}.
 \label{eq:delta_f}
\ee
Further, we evaluate the resonance (ILR) contribution into the sum over $l$, hence we consider the term corresponding to $l=-1$:
\be
\delta f^{\rm RES}=\frac{e^{2i(\varphi-\Omega_{\rm p}t)}}{\varkappa^2\,a}
\frac{{\p F}/{\p a}}
{\nu(R)+1}\, \phi_{-1}(R,a)\,e^{-iw_1}.
\label{eq:delta_f_res}
\ee
First, we calculate $\phi_{-1}(R,a)$. By definition
$$
\phi_l(I_1,I_2)=\frac{1}{2\pi} \int\limits_{-\pi}^{\pi} V(r)\,
\exp\Bigl\{i\,\Bigl[\int\limits^r k(r')\,\d r'-lw_1\Bigr]\Bigr\}\,\d w_1.
$$
In the leading order, one can replace $r$ by $R$ in amplitude $V(r)$. In the exponent, however, due to the large wavenumber $k$, we have to retain the upper limit $r$ in the integral $\int^r k(r')\,\d r'$, so
\be
\int\limits^r k(r')\,\d r'=\int\limits^{R-a\cos w_1} k(r')\,\d r'\approx
\int\limits^{R} k(r')\,\d r' - ka\cos w_1.
 \label{eq:int_k}
\ee
Then we find
\begin{multline}
\phi_{-1}(I_1,I_2)=\frac{1}{2\pi}\,V(R)\,\exp\Bigl[i\!\int\limits^R k(r')\,\d r'\Bigr]\\ \times\int\limits_{-\pi}^{\pi}
e^{-i(ka\cos w_1+lw_1)}\,\d w_1.
 \label{eq:phi-1}
\end{multline}
Calculating the integral over $w_1$ in r.h.s. of (\ref{eq:phi-1}) \citep[][]{GR15}, we finally obtain:
\begin{multline}
\delta f^{\rm RES}=-i\,V(R)\,\frac{e^{2i(\varphi-\Omega_{\rm p}t)}}{\varkappa^2\,a}
\frac{{\p F}/{\p a}}
{\nu(R)+1}\, e^{-iw_1}\,\\
\times\exp\Bigl[i\!\int\limits^R k(r')\,\d r'\Bigr]\,J_1(ka),
 \label{eq:fres_1}
\end{multline}
where $J_n(z)$ is the Bessel function. In the perturbed surface density $\delta\sigma$
$$
\delta\sigma(r,\varphi,t)=S(r)\,\exp\Bigl\{i\,\Bigl[\int\limits^r k(r')\,\d r'+2\varphi-2\Omega_{\rm p} t\Bigr]\Bigr\}
$$
evaluation of $S^{\rm RES}$ using (\ref{eq:fres_1}) and (\ref{eq:int_k}) for the integral $\int^R k(r')\,\d r'$ gives:
\be
S^{\rm RES}=-i \frac{2\,V}{\varkappa^2}\int
\frac{\p F/\p (a^2)\,J_1(ka)}
{\nu(R)+1}\, e^{ika\cos w_1-iw_1} \d\Gamma.
 \label{eq:Sres}
\ee
{
It is easy to see that in nonresonant case (\ref{eq:Sres}) gives the known expression for corresponding $(l=-1)$ contribution into the density response. In this case denominator  $\nu(R)+1$
can be replaced by $\nu(r)+1$ and then taken out of the integral. The volume element in velocity space
$\d\Gamma=\d v_r \d v_{\varphi}$ can be written (see \ref{eq:dGamma}) as
$\d\Gamma={\varkappa^3(r)}/[2\Omega(r)]\,a\,\d a\,\d w_1,
$
and the equilibrium DF as
\be
F=\frac{2\Omega(r)}{\kappa(r)}\,\frac{\sigma(r)}{2\pi c^2(r)}\,\exp\Bigl[-\frac{\varkappa^2(r)\,a^2}{2\,c^2(r)}\Bigr].
  \label{eq:F_a}
\ee
Using the expression for the derivative $\p F/\p(a^2)$:
\be
\frac{\p F}{\p (a^2)}=
-\frac{2\Omega(r)}{\varkappa(r)}\,\frac{\sigma(r)\,\varkappa^2(r)}{4\pi c^4(r)}
\exp\Bigl[-\frac{\varkappa^2(r)\,a^2}{2\,c^2(r)}\Bigr],
  \label{eq:dF_da}
\ee
we find
\begin{multline}
S^{\rm RES}(r)=i \frac{V(r)}{\varkappa^2(r)}\,\left(\frac{\varkappa}{c}\right)^4\,
\frac{\sigma_0(r)}{\nu+1}
\int\limits_0^{\infty}
e^{-(\varkappa\,a/c)^2/2}\,a\,\d a\,J_1(ka)\\
\times\int\limits_{-\pi}^{\pi} e^{ika\cos w_1-iw_1}\,\frac{\d w_1}{2\pi}.
\end{multline}
Integration over $w_1$ yields
$$\int\limits_{-\pi}^{\pi} e^{ika\cos w_1-iw_1}\,\frac{\d w_1}{2\pi}=i\,J_1(ka),
$$
so that
$$
S^{\rm RES}=
-\frac{V}{\varkappa^2}\left(\frac{\varkappa}{c}\right)^4
\frac{\sigma_0}{\nu+1}
\int\limits_0^{\infty}\!
e^{-(\varkappa\,a/c)^2/2}\,a\,\d a\,[J_1(ka)]^2.
$$

The integration over $a$ can be done easily \citep[see eq. (663.2) in][]{GR15}, and in the nonresonant limit for  $l=-1$ one can have:
\be
S^{\rm RES}=-\frac{V\,\sigma_0}{(\varkappa\, r)^2\epsilon^2}\,\frac{I_1(x)\,e^{-x}}{\nu(r)+1},
 \label{eq:S_LS}
\ee
where
$x=(kr\epsilon)^2=(k c/\varkappa)^2$, $\epsilon = c/(r\varkappa)$, $I_n(z)$ is the modified Bessel function,
in accordance with the known result by \cite{LS66}.
\medskip

Let's go back to the expression (\ref{eq:Sres}) for $S^{\rm RES}$. Near the resonance we have
\be
\nu(R)+1=\nu'(R_L)(R-R_L).
\label{eq:nu+1}
\ee
\cite{Mark71, Mark74} introduces an auxiliary scale $\ell$ instead of (\ref{eq:nu+1}),
\be
\nu(R)+1=\frac{R-R_L}{\ell},
\label{eq:nu+1_mark}
\ee
taking into account that for real galaxies $\nu'(R_L)>0$ ($\ell^{-1}$ does not coincide with $\nu'(R_L)$), in order to extend the linear law for $\nu (R)+1$ versus $R-R_L$  to the largest possible neighbourhood of the resonance, including the region where this linear law is no longer valid. In other words, instead of the true slope of the curve $\nu(R)$ at $R=R_L$, he uses an average slope, $\nu(R) +1\approx (R-R_L)/\ell$. Substituting $R=r+a\,\cos w_1$ in (\ref{eq:nu+1_mark})
$$\nu(R)+1\approx
({r}/{\ell})\, (\rho\cos w_1-\eta_L),
$$
where $\eta_L=-(r-R_L)/{r}$, $\rho={a}/{r}$, we have
\begin{multline}
S^{\rm RES}(r)=-i\, \frac{\ell}{r}\,\frac{2\,V(r)}{\varkappa^2(r)}\cdot \dfrac{\varkappa^3(r)}{2\Omega(r)}
\int\limits_0^{\infty} a\,\d a\\ \times\int\limits_{-\pi}^{\pi} \d w_1
\frac{\p F/\p (a^2)\,J_1(ka)}
{\rho\cos w_1-\eta_L}\, e^{ika\cos w_1-iw_1}.
 \label{eq:Sres_a}
\end{multline}
In what follows we should distinguish two possible version of bypass rule -- {\it standard} and {\it modified}.

\bigskip
\centerline{\it The standard bypass rule}
\medskip

In order to proceed further on the basis of the causality principle one needs to calculate the integral
(\ref{eq:Sres_a}) with the the pole using the {\it standard} Landau\,-\,Lin bypass rule. This can be done by adding a small imaginary {\it positive} part $\omega\to \omega+i 0^+$ to the frequency. Since $\nu(R)=2[\omega-\Omega(R)]/\varkappa(R)$, it is equivalent to assigning a small imaginary part to $\nu(R)$.
Then, provided that $\ell\approx 1/\nu'(R_L)$ is positive,
$$
\nu(R)+1\to \nu(R)+1+i\,0^+ =
({r}/{\ell})\, [\rho\cos w_1-(\eta_L-i\,0^+)].
$$
In terms of variable $\eta=a\,\cos w_1$, the resonance denominator is
$
({r}/{\ell})\, [\eta-(\eta_L-i\,0^+)]$. It means the bypass of the pole in complex $\eta$ plane from above.
The direction of the bypass, of course, depends on the sign of the derivative $\nu'(R_L)$. In our case, it corresponds to the positive sign $\nu'$.

If we  express the resonant denominator as
\begin{multline}
\frac{1}{\rho\cos w_1-\eta_L}\to \frac{1}{\rho\cos w_1-\eta_L+i 0}\\
=-i\int\limits_{-\infty}^0 \d\lambda\, e^{-i(\rho\cos w_1-\eta_L+i 0)\,\lambda},
\label{eq:denominator}
\end{multline}
then for the integral over $w_1$ in the r.h.s. of (\ref{eq:Sres_a}), we find
\begin{multline}
\int_{-\pi}^{\pi} \, \frac{e^{ika\cos w_1-iw_1}}{\rho\cos w_1-\eta_L}\,\d w_1\\=
2\pi\!\!\int\limits_{-\infty}^0 \!\!\d\lambda\, e^{i\eta_L\lambda}\,J_1[(a/r)(kr-\lambda)].
 \label{eq: int_dw}
\end{multline}
Substituting  (\ref{eq: int_dw}) into (\ref{eq:Sres_a}) and using (\ref{eq:dF_da}) for $\p F/\p a^2$ we have
\begin{multline}
S^{\rm RES}(r)=
i\,\frac{\ell}{r}\, V(r)\,\sigma_0(r)\,\frac{\varkappa^2(r)}{c^4(r)}
\int\limits_{-\infty}^0 \d\lambda\, e^{i\eta_L\lambda}\\ \times \int\limits_0^{\infty}
\exp\Bigl[-\frac{\varkappa^2(r)\,a^2}{2\,c^2(r)}\Bigr]\,
J_1[(a/r)(kr-\lambda)]\,J_1(ka)\,a\, \d a.
 \label{eq: Sres_b}
\end{multline}
For the integral over $a$ in r.h.s. of (\ref{eq: Sres_b}) \citep[see eq. (663.2) in][]{GR15} we find
\begin{multline}
\int\limits_0^{\infty}
\exp\Bigl[-\frac{\varkappa^2(r)\,a^2}{2\,c^2(r)}\Bigr]\,
J_1[(a/r)(kr-\lambda)]\,J_1(ka)\,a\, \d a\\
=
\frac{c^2(r)}{\varkappa^2(r)}\, \exp\Bigl[-\Bigl(x-y\,\frac{c\lambda}{\varkappa\, r}\Bigr)\Bigr]\,I_1\Bigl(x-y\,\frac{c\lambda}{\varkappa\, r}\Bigr)\,
e^{-(\lambda c/\varkappa r)^2/2}.
\end{multline}
Denoting
$
\mu=({c}/{\varkappa\,r})\,\lambda\equiv \epsilon\,\lambda
$
and
$-{\eta_L}/{\epsilon}\equiv(r-R_L)/(r\,\epsilon)=u_L,
$
we find the final expression for the resonance surface density response for the bypass from above:
\begin{multline}
 S^{\rm RES}(r)=i\,\ell\,\frac{V\,\sigma_0}{r\,(\varkappa\,
 r)^2\epsilon^3}\\ \times
\int\limits_{-\infty}^0 \d\mu\,e^{-i\mu u_L}e^{-\mu^2/2}
I_1(x-\mu\,y)\,e^{-(x-\mu\,y)},
  \label{eq:Sres_LL}
\end{multline}
where
$
y=(kr)\,\epsilon={kc}/{\varkappa}, \ \ x=y^2.
$
The reduction factor can be found from the relation
\be
S^{\rm RES}(r)=-\frac{k^2(r)}{\varkappa^2(r)}\,V(r)\,\sigma_0(r)\,{\cal F}^{\rm RES},
 \label{eq: S-RF}
\ee
that yields
\begin{multline}
{\cal F}^{\rm RES}_{\frown}(y,u_L)= -\frac{i\,\ell}{r\,\epsilon\,x}\\
\times\int\limits_{-\infty}^0\,I_1(x-\mu\,y)\,\,e^{-(x-\mu\,y)}
\,e^{-\frac{1}{2}\,\mu^2-i\,u_L\,\mu}\,\d\mu.
\label{eq: RF_Mark}
\end{multline}
The symbol `$\frown$' means bypass of the pole in complex plane from above. The variable $u_L=(r-R_L)/(r\epsilon)$ characterises the distance to resonance circle measured in units of epicyclic size $a = r\epsilon=c/\varkappa$.

For further simplifications it is convenient to rewrite (\ref{eq: RF_Mark}) in an equivalent form using
integral presentation (\ref{eq:bessel}) for $I_1(z)\,\exp(-z)$. We obtain
\be
{\cal F}^{\rm RES}_{\frown}=\frac{\ell}{r\,\epsilon\,x}\,F_{\frown}^{\rm RES},
\ee
where
\begin{multline}
F_{\frown}^{\rm RES}
=-i\int\limits_{-\infty}^0\,I_1(x-\mu\,y)\,\,e^{-(x-\mu\,y)}
\,e^{-\frac{1}{2}\,\mu^2-i\,u_L\,\mu}\,\d\mu\\
=\frac{1}{\pi}\int\limits_0^{\pi} e^{-y^2\,(1-\cos\theta)}\cos\theta\,\d\theta\\
\times \Bigl[e^{-Z^2/2}\Bigl(\int\limits_0^Z e^{s^2/2} \d s-i\,\sqrt{\frac{\pi}{2}}\Bigr)\Bigr],
\label{eq:F_frown}
\end{multline}
and $Z=u_L+iy\,(1-\cos\theta)$.

\bigskip
\centerline{\it The modified bypass rule}
\medskip

In the case of modified bypass rule we have to take instead of (\ref{eq:denominator})
the half-sum of the expressions corresponding to bypass from above and from below:
\begin{multline}
\frac{1}{\rho\cos w_1-\eta_L}\to \\ \to \frac{1}{2}\left(\frac{1}{\rho\cos w_1-\eta_L+i 0}
+ \frac{1}{\rho\cos w_1-\eta_L-i 0}\right) \\
=-\frac{i}{2}\,\left(\int\limits_{-\infty}^0+\int\limits_{+\infty}^0\right) \d\lambda\, e^{-i(\rho\cos w_1-\eta_L)\,\lambda}.
\label{eq:denominator_mod}
\end{multline}
This leads to modification of the expression for the surface density response:
\begin{multline}
 S^{\rm RES}(r)=\frac{i\,\ell}{2}\,\frac{V\,\sigma_0}{r\,(\varkappa\,
 r)^2\epsilon^3}\\ \times
\left(\int\limits_{-\infty}^0+\int\limits_{+\infty}^0\right) \d\mu\,e^{-i\mu u_L}e^{-\mu^2/2}
I_1(x-\mu\,y)\,e^{-(x-\mu\,y)}.
  \label{eq:Sres_mod}
\end{multline}
The modified reduction factor is
\begin{multline}
{\cal F}^{\rm RES}_P(y,u_L)= -\frac{i\,\ell}{2\,r\,\epsilon\,x}\\
\times \left(\int\limits_{-\infty}^0+\int\limits_{+\infty}^0\right)\,
I_1(x-\mu\,y)\,\,e^{-(x-\mu\,y)}\,e^{-\frac{1}{2}\,\mu^2-i\,u_L\,\mu}\,\d\mu.
\end{multline}
Here index `$P$' means the integral calculated in the principal value sense. The equivalent form for ${\cal F}^{\rm RES}_P$, obtained with the help of  (\ref{eq:bessel}) is
\be
{\cal F}_P^{\rm RES}=\frac{\ell}{r\,\epsilon x}\,F_P^{\rm RES},\ \ \ F_P^{\rm RES}=\frac{F^{\rm RES}_{\frown}+F^{\rm RES}_{\smile}}{2},
\label{eq:F_P_RES}
\ee
where
\begin{multline}
F^{\rm RES}_{\smile}=-i\int\limits_{+\infty}^0
I_1(x-\mu\,y)\,\,e^{-(x-\mu\,y)}\,e^{-\frac{1}{2}\,\mu^2-i\,u_L\,\mu}\,\d\mu\\
=
\frac{1}{\pi}\int\limits_0^{\pi} e^{-y^2\,(1-\cos\theta)}\cos\theta \,\d\theta\\
\times\Bigl[e^{-Z^2/2}\Bigl(\int\limits_0^Z e^{s^2/2} \d s+i\,\sqrt{\frac{\pi}{2}}\Bigr)\Bigr],
\label{eq:F_smile}
\end{multline}
and symbol '$\smile'$ means bypass from below. Thus, taking the half-sum of (\ref{eq:F_frown}) and
(\ref{eq:F_smile})  we find
\be
F^{\rm RES}_P
=\frac{1}{\pi}\int\limits_0^{\pi} e^{-y^2\,(1-\cos\theta)}\cos\theta \,\d\theta\cdot
\Bigl(e^{-Z^2/2}\int\limits_0^Z e^{s^2/2} \d s\Bigr).
\label{eq:F_P}
\ee

\section{Asymptotic forms of the dispersion relations}

\subsection{The standard bypass (from above)}

Substitution of ${\cal F}^{\rm RES}_{\frown}$ into r.h.s. of (\ref{eq:stand1}) yields the dispersion relation (without the nonresonant terms) for the case of the bypass from above \citep[see also][]{Mark71, Mark74}:
\be
y^2=y_L^2 (2s_y y)\,F^{\rm RES}_{\frown}(y,u_L),
\label{eq:DE_Mark}
\ee
where function $F_{\frown}^{\rm RES}(y,u_L)$ is defined by eq. (\ref{eq:F_frown}) and
 $s_y={\rm sgn}[{\rm Re}(y)]=s_k$,  $y_T=k_T\,(r\epsilon)$, $y_L^2=k_L^2\,(r\epsilon)^2$,
$k_L^2= \pi G\sigma_0\varkappa^2 \ell/{c^4}$.
Our goal is to simplify $F_{\frown}^{\rm RES}(y,u_L)$ for the large $x_r={\rm Re}(x)\equiv {\rm Re}(y^2)$.

For $x_r\gg 1$, the main contribution to the integral over $\theta$ comes from two narrow regions near: (i) $\theta \approx 0$ and (ii) $\theta\approx \pi$. The  width of these regions is $\Delta\theta\sim x_r^{-1/2}$, as is clear from the expression (\ref{eq:F_frown}), which can be written in the form:
\begin{multline}
F_{\frown}^{\rm RES}(y,u_L)=\frac{1}{\pi}\int\limits_0^{\pi} e^{-\frac{1}{2}\,y^2\sin^2\theta-i u_L y\,(1-\cos\theta)} \cos\theta\,\d\theta\\ \times \Biggl[\int\limits_0^{u_L+i\,y\,(1-\cos\theta)}e^{s^2/2}\,\d\mu-
i\,\sqrt{\frac{\pi}{2}}\Biggr].
 \label{F_frown1}
\end{multline}
Let us calculate these contributions separately denoting them $F_{\frown}^{(1)}$ and $F_{\frown}^{(2)}$, respectively.
\medskip

(i) For the region  $\theta\ll 1$ we have:
\be
F_{\frown}^{(1)}(y,u_L)=\sqrt{\frac{1}{2\pi}}\,
\frac{s_y}{y}\,e^{-u_L^2/2}\,\Biggl(-i\,\sqrt{\frac{\pi}{2}}+\int\limits_0^{u_L} \d s\,e^{s^2/2}\Bigr),
\ee
where $s_y={\rm sgn}(y)={\rm sgn} (k)$.
\medskip

(ii) For the region $\pi-\theta\ll 1$ we have:
\be
F_{\frown}^{(2)}(y,u_L)=-i\,\sqrt{\frac{1}{2\pi}}\,
\frac{s_y}{y}\,\sqrt{\frac{\pi}{2}}\,(s_y-1)\,e^{-2iu_Ly-u_L^2/2}.
\ee
Summarising both contributions,
\begin{multline}
F^{\rm RES}_{\frown}(y,u_L) \approx F_{\frown}^{(1)}(y,u_L)+F_{\frown}^{(2)}(y,u_L)\\
=\frac{s_y}{2y}\,e^{-u_L^2/2}\,\Biggl\{-i\,\Bigl[1+(s_y-1)\,e^{-2iu_L y}\Bigr]+
\sqrt{\frac{2}{\pi}}\int\limits_0^{u_L} \d s\,e^{s^2/2}\Biggr\}.
\label{eq:F_frown_asymp}
\end{multline}

Finally the approximate dispersion relation for the case of bypass from above becomes
\be
y^2=y_L^2\,e^{-u_L^2/2}\,\Biggl\{-i\,\Bigl[1+(s_y-1)\,e^{-2iu_{\!L} y}\Bigr]+
\sqrt{\frac{2}{\pi}}\int\limits_0^{u_{\!L}} \d s\,e^{s^2/2}\Biggr\}.
\label{eq:DE_stand_asymp}
\ee

\subsection{The modified bypass rule}
In this case we have instead of (\ref{eq:DE_Mark})
\be
y^2=y_L^2 (2 y s_y)\,F^{\rm RES}_P(y,u_L),
\ee
where $F^{\rm RES}_P$  is defined by eqs. (\ref{eq:F_P_RES}) and (\ref{eq:F_P}).
Calculations  analogous to those performed for the asymptotic form of $F_{\frown}^{\rm RES}(y,u_L)$ yield for $F_{\smile}^{\rm RES}(y,u_L)$
\begin{multline}
F_{\smile}^{\rm RES}(y,u_L)=F_{\smile}^{(1)}(y,u_L)+F_{\smile}^{(2)}(y,u_L)\\
=\frac{s_y}{2y}\,e^{-u_L^2/2}\,\Biggl\{i\,\Bigl[1-(s_y+1)\,e^{-2iu_L y}\Bigr]+
\sqrt{\frac{2}{\pi}}\int\limits_0^{u_L} \d s\,e^{s^2/2}\Biggr\}.
\label{eq:F_smile_asymp}
\end{multline}
Summarising (\ref{eq:F_frown_asymp}) and (\ref{eq:F_smile_asymp}),
\begin{multline}
F^{\rm RES}_P(y,u_L)=\frac{1}{2}\,\Bigl(F_{\frown}^{\rm RES}+F_{\smile}^{\rm RES}\Bigr)\\
\approx \frac{s_y}{2y}\,e^{-u_L^2/2}\,\Biggl(-i\,s_y\,e^{-2iu_L y}+
\sqrt{\frac{2}{\pi}}\int\limits_0^{u_L} \d s\,e^{s^2/2}\Biggr).
\label{F_P_asymp)}
\end{multline}

Finally, for the modified bypass rule we obtain the following approximate dispersion equation instead of
(\ref{eq:DE_stand_asymp}):
\be
y^2=y_L^2\,e^{-u_L^2/2}\,\Biggl(-i\,s_y\,e^{-2\,iu_{\!L} y}+
\sqrt{\frac{2}{\pi}}\int\limits_0^{u_{\!L}} \d s\,e^{s^2/2}\Biggr).
 \label{eq:DE_modif_asymp}
\ee


\begin{thebibliography}{99}

\bibitem[\protect\citeauthoryear{Benney \& Bergeron}{1969}]{BB69}   Benney D. and  Bergeron R.E., 1969, Studies in Appl. Math.,  48, 181

\bibitem[\protect\citeauthoryear{Bertin}{2014}]{B14}  Bertin G., 2014,  Dynamics of Galaxies: 2nd ed. (Cambridge Univ. Press, Cambridge, UK, 2014).

\bibitem[\protect\citeauthoryear{Binney \& Tremaine}{2008}]{BT2008}  Binney J. and  Tremaine S., 2008,  Galactic Dynamics: 2nd ed. (Princeton Univ. Press, NJ, 2008).

\bibitem[\protect\citeauthoryear{Churilov \& Shukhman}{1996\,a}]{CS96a}  Churilov S.M. and Shukhman I.G., 1996a, Atmospheric and Oceanic Physics, 31, No 4, 535

\bibitem[\protect\citeauthoryear{Churilov \& Shukhman}{1996b}]{CS96b}  Churilov S.M. and Shukhman I.G., 1996b, J. Fluid Mech., 318,  189

\bibitem[\protect\citeauthoryear{Combes \& Elmegreen}{1993}]{CE93} Combes F. and  Elmegreen B.G., 1993,  A\&A,   271, 31

\bibitem[\protect\citeauthoryear{Contopoulos}{1970a}]{Contopoulos70a}  Contopoulos G., 1970a, Proc. IAU Symposium No 38,
Eds.: Becker W. and  Contopoulos G.I., P. 303

\bibitem[\protect\citeauthoryear{Contopoulos}{1970b}]{Contopoulos70b}  Contopoulos G., 1970b, ApJ, 160, 113

\bibitem[\protect\citeauthoryear{Contopoulos}{1971}]{Contopoulos71}  Contopoulos G., 1971, ApJ, 163, 181

\bibitem[\protect\citeauthoryear{Davis}{1969}]{Davis69} Davis R.E., 1969, J. Fluid Mech.,  36, 337

\bibitem[\protect\citeauthoryear{Dickinson}{1970}]{Dick70}  Dickinson R.E., 1970,  J. Atmos. Sci.,  27, 627

\bibitem[\protect\citeauthoryear{Dobbs \& Baba}{2014}]{DB14} Dobbs C. and Baba J., 2014, Publications of the Astronomical Society of Australia, 31, 1

\bibitem[\protect\citeauthoryear{Goldreich \& Tremaine}{1981}]{GT81} P. Goldreich P. and  Tremaine S., 1981, ApJ,   243, 1062

\bibitem[\protect\citeauthoryear{Gradshteyn \& Ryzhik}{2015}]{GR15} Gradshteyn I. S.,  Ryzhik I. M., 2015, Table of Integrals, Series, and Products. Edited by Zwillinger D. and  Moll V.,  Academic Press, New York, 8th edition

\bibitem[\protect\citeauthoryear{Haberman}{1972}]{Haberman72}  Haberman R., 1972, Studies in Appl. Math.,  51, 139


\bibitem[\protect\citeauthoryear{Killworth \& McIntyre}{1985}]{KM85}   Killworth P.D. and  McIntyre M.E., 1985, J. Fluid Mech.,  161, 449


\bibitem[\protect\citeauthoryear{Landau}{1946}]{Landau46}  Landau L.D., 1946,  J. Phys (USSR),  10, 25 (Reproduced in Uspekhi Fizicheskih Nauk, 1967, 93 (11), 527, in Russian)

\bibitem[\protect\citeauthoryear{Lau}{1976}]{Lau76} Lau Y.Y., 1976, Physics of Fluids, 19, 1644

\bibitem[\protect\citeauthoryear{Lau \& Bertin}{1978}]{LauBertin78} Lau Y.Y. and Bertin G., 1978, ApJ, 226, 508

\bibitem[\protect\citeauthoryear{Lau et al.}{1976}]{LauLinMark76} Lau Y.Y., Lin C.C. and Mark J. W.-K., 1976, Proc. Nat.  Acad.  Sci. USA,  73, 1379

\bibitem[\protect\citeauthoryear{Lin}{1955}]{Lin55}  Lin C.C., 1955, The Theory of Hydrodynamic Stability (Cambridge University Press, New York, 1955)

\bibitem[\protect\citeauthoryear{Lin}{1970}]{Lin70}  Lin C.C., 1970, Proc. IAU Symposium No 38,
Eds.: Becker W. and  Contopoulos G.I., P. 377

\bibitem[\protect\citeauthoryear{Lin \& Shu}{1966}]{LS66}   Lin C.C. and  Shu F., 1966, Proc. Nat.  Acad.  Sci. USA,  55, 229

\bibitem[\protect\citeauthoryear{Lin \& Lau}{1975}]{LinLau75}   Lin C.C. and  Lau Y.Y., 1975, SIAM J. Appl. Math.,  29, 352

\bibitem[\protect\citeauthoryear{Lin \& Lau}{1979}]{LinLau79}   Lin C.C. and  Lau Y.Y., 1979, SIAM J. Appl. Math.,  60, 97

\bibitem[\protect\citeauthoryear{Lyakhovich et al.}{1994}]{LFK94}   Lyakhovich V.V., Fridman A.M., Khoruzhii O.V., 1994,  In {\it ``Unstable processes in Universe''}, Ed.: Masevich A.G., Moscow, Cosmosinform, P. 194, in Russian

\bibitem[\protect\citeauthoryear{Lynden-Bell \& Kalnajs}{1972}]{LBK72} Lynden-Bell D. and Kalnajs A.J.,  1972, MNRAS,  157, 1 (LBK)

\bibitem[\protect\citeauthoryear{Lynden-Bell \& Ostriker}{1967}]{LBO67} Lynden-Bell D. and Ostriker J.P.  1967, MNRAS,  136, 293

\bibitem[\protect\citeauthoryear{Mark}{1971}]{Mark71} Mark J. W-K., 1971, Proc. Nat.  Acad.  Sci. USA,  68, 2095

\bibitem[\protect\citeauthoryear{Mark}{1974}]{Mark74} Mark J. W-K., 1974, ApJ, 193, 339

\bibitem[\protect\citeauthoryear{Mark}{1976}]{Mark76} Mark J. W-K., 1976, ApJ, 203, 81

\bibitem[\protect\citeauthoryear{Mark}{1977}]{Mark77} Mark J. W-K., 1977, ApJ, 212, 645

\bibitem[\protect\citeauthoryear{Maslowe}{1986}]{Maslowe86}  Maslowe S.A., 1986, Ann. Rev. Fluid Mech., 18, 405

\bibitem[\protect\citeauthoryear{Pannatoni}{1983}]{Pannatoni1983} Pannatoni R.F., 1983, Geophys. \& Astrophys. Fluid Dynamics, 24, 165

\bibitem[\protect\citeauthoryear{Polyachenko}{2013}]{Pol2013} Polyachenko E.V., 2013, Astron. Lett., 39, 72

\bibitem[\protect\citeauthoryear{Polyachenko et al.}{2016}]{PBJ2016}  Polyachenko E.,  Berczik P., and  Just A, 2016, MNRAS,  462, 3727

\bibitem[\protect\citeauthoryear{Polyachenko}{2018}]{Pol2018} Polyachenko E.V., 2018, MNRAS,  478, 4268

\bibitem[\protect\citeauthoryear{Shu}{1970}]{Shu70}  Shu F.H., 1970, ApJ,  160, No 1, 99

\bibitem[\protect\citeauthoryear{Stewartson}{1978}]{Stew78} Stewartson K., 1978,  Geophys. Astrophys. Fluid Dynamics, 9, 185

\bibitem[\protect\citeauthoryear{Toomre}{1964}]{Toomre64}   Toomre A., 1964, ApJ,   139, 1217

\bibitem[\protect\citeauthoryear{Toomre}{1977}]{T77} Toomre A., 1977, ARA\&A, 15, 437

\bibitem[\protect\citeauthoryear{Toomre}{1981}]{Toomre81}  Toomre A., 1981,  {\it in Structure and evolution of normal galaxies}. Eds.: Fall S.M.,  Lynden-Bell D., Cambridge  Univ. Press, P. 111

\bibitem[\protect\citeauthoryear{Vandervoort}{1973}]{Vandervoort73}   Vandervoort P.O., 1973, ApJ,   180, 739

\bibitem[\protect\citeauthoryear{Vandervoort}{1975}]{Vandervoort75}   Vandervoort P.O., 1975, ApJ,   201, 50

\bibitem[\protect\citeauthoryear{Vandervoort \& Monet}{1975}]{VM75}   Vandervoort P.O. and Monet D.G., 1975, ApJ,   201, 311

\bibitem[\protect\citeauthoryear{Vlasov}{1945}]{Vlasov1945}  Vlasov A.A., 1945, J. Phys. (USSR),  9, 25

\bibitem[\protect\citeauthoryear{Warn \& Warn}{1978}]{WW78}  Warn T. and H. Warn H., 1978,  Studies in Appl. Math.,  59, 37

\end{thebibliography}
\end{document}